\documentclass[prd,eqsecnum,showpacs,nofootinbib,preprintnumbers]{revtex4}

\usepackage{amsmath,amssymb,amsfonts}
\usepackage[dvips]{graphicx,color}

\newcommand{\ma}[1]{{\mathrm{#1}}}
\newcommand{\pa}{{\partial}}

\begin{document}
\preprint{YITP-11-3}
\title{Conservation of the nonlinear curvature perturbation
 in generic single-field inflation}

\author{Atsushi Naruko and Misao Sasaki}

\affiliation{Yukawa Institute for Theoretical Physics, Kyoto University,
 Kyoto 606-8502, Japan
}

\date{\today}

\begin{abstract}
It is known that the curvature perturbation on uniform energy density
 (or comoving or uniform Hubble ) slices on superhorizon scales 
is conserved to full nonlinear 
order if the pressure is only a function of the energy
density (ie, if the perturbation is purely adiabatic),
independent of the gravitational theory. 
Here we explicitly show that the same conservation holds
for a universe dominated by a single scalar field provided that
the field is in an attractor regime, for a very general class of scalar 
field theories. However, we also show that if the scalar field equation
 contains a second time derivative of the metric, as in the case of 
the Galileon (or kinetic braiding) theory,
one has to invoke the gravitational field equations
to show the conservation.
\end{abstract}


\maketitle

\section{introduction}

The CMB anisotropy observation by WMAP clearly showed us that
the primordial curvature perturbations are nearly scale invariant
and their statistics is Gaussian to very high accuracy~\cite{wmap-five}.
This is perfectly in agreement with the predictions of the standard,
canonical single-field, slow-roll inflation. 
Nevertheless, an indication was found in the CMB data as well as in the
large scale structure data that there may be a detectable level of
non-Gaussianity in the curvature perturbation.
Consequently a lot of attention has been paid in recent years to possible
non-Gaussian features in the primordial curvature perturbation from
inflation \cite{komatsu-spergel}. (See articles in a focus section
in CQG~\cite{sasaki-wands} and references therein for recent
developments.)
Apparently this direction of research involves nonlinear
cosmological perturbations.

There are mainly two approaches to nonlinear cosmological perturbations.
One is the standard perturbative approach
 \cite{maldacena,malik-wands,seery-lidsey,not-so-short}.
This is basically straightforward and can in principle deal with most
general situations as long as the perturbation expansion is applicable.
But the equations can become
very much involved and quite often the physical transparency may be lost.

The other is the gradient expansion approach
 \cite{russia-lifshitz,russia-belinsky,starobinsky,bardeen,salopek-bond,
nathalie-langlois,sasaki-stewart,nambu-taruya,sasaki-tanaka,shibata-sasaki,
lyth-rodriguez,rigopoulos-shellard1,rigopoulos-shellard2,yokoyama1,yokoyama2,
yotanaka1,yotanaka2,takamizu-mukohyama,takamizu-sasaki,LMS,afshordi}. 
In this approach, the field equations are expanded
 in powers of spatial gradients.
Hence this is applicable only to perturbations on superhorizon scales.
Nevertheless, it has a big advantage that the full nonlinear effects
are taken into account at each order of the gradient expansion.
At leading order in gradient expansion, it corresponds to the
separate universe approach \cite{separate-universe}. 
Namely, the field equations become ordinary 
differential equations with respect to time, hence the physical quantities 
at each spatial point (where 'each point' corresponds to a Hubble horizon
size region) evolve in time independently from those at
the rest of the space.

One of the most important results obtained in the gradient expansion
approach is that the full nonlinear curvature perturbation
on unifrom energy density (or comoving or uniform Hubble) slices
is conserved at leading order in gradient expansion
if the pressure is only a function of the energy density~\cite{LMS},
or the perturbation is purely adiabatic. This is shown using the
energy conservation law, without using the Einstein equations.
Thus, without solving the field equations,
one can predict the spectrum and the statistics 
of the curvature perturbation at horizon re-entry during 
the late radiation or matter-dominated era once one knows these 
properties of the curvature perturbaiton at horizon exit during inflation.

However, the assumption that the pressure is only a function of
the energy density is not rigorously true in the case of
a scalar field. It is only approximately true in the limit of the
slow-roll inflation. In other words, the pressure and energy density
perturbations are not `adiabatic' in the sense of the stardard
fluid dynamics.
Therefore it is not completely clear exactly under which
condition the nonlinear curvature perturbation is
conserved in the case of a scalar field.

In this short note, we focus on a universe dominated by a
single scalar field $\phi$ and explicitly show the conservation of 
the nonlinear curvature perturbation on comoving slices 
$(\phi = \phi (t))$ at leading order in gradient expansion.
We consider a very general theory of a scalar field \cite{k-inflation,
garriga-mukhanov,DBI,DBI-in-the-sky}, including a Galileon 
(or kinetic braiding) field which has been attracting attention
recently \cite{galileon,g-inflation,galileon-mizuno,galileon-deffayet,KMD}.
For the gravitational part we assume Einstein gravity for definiteness,
but our discussion is applicable to any metric theory of gravity.

Assuming that the scalar field dynamics is in an attractor regime
so that the value of the scalar field determines the dynamics
completely, we find that the conservation of the nonlinear curvature
perturbation holds without using the gravitational field
equations just as the same as the fluid case,
provided that the scalar field equation contains only first
time derivatives of the metric. This condition is satisfied
for a generic K-essence type scalar field, including the
case of a canonical scalar, but not for a Galileon scalar
because the Galileon field equation contains second time
derivatives of the metric. In the latter case, one
has to invoke the gravitational field equations to see
if the conservation still holds or not. In the case of
Einstein gravity, the conservation is shown to hold
even for a Galileon scalar field.

\section{basic setup}
We focus on the dynamics on superhorizon scales. We associate
$\epsilon$ to each spatial derivative. So a quantity with
$n$-th spatial derivatives will be of $O(\epsilon^n)$.

We express the metric in the $(3+1)$ form,
\begin{gather}
 ds^2 =g_{\mu\nu}dx^\mu dx^\nu
= - \alpha^2 dt^2
 + \gamma_{i j} (dx^i + \beta^i dt) (dx^j + \beta^j dt)\,,
\end{gather}
where $\alpha$, $\beta^i$ and $\gamma_{i j}$ are the lapse function,
the shift vector, and the spatial metric, respectively.
We choose the spatial coordinates such that $\beta^i=O(\epsilon^3)$.
We further decompose the spatial metric as
\begin{gather}
 \gamma_{i j} = a^2 (t) e^{2 \psi(t,x^k)} \tilde{\gamma}_{i j}(t,x^k)\,,
\quad \det \tilde{\gamma}_{i j} = 1\,,
\end{gather}
where $a(t)e^{\psi(t,x^k)}$ is the scale factor at each local point
while $a(t)$ is the scale factor of a fiducial homogeneous universe.
When the gradient expansion is applied to an inflationary
stage of the universe, it is known that we have
$\pa_t \tilde{\gamma}_{i j} = O (\epsilon^2)$ \cite{LMS,yotanaka1,yotanaka2}.
Then at leading order in gradient expansion,
we identify $\psi$ as the nonlinear curvature perturbation~\cite{LMS}.

We consider a theory with the action,
\begin{eqnarray}
 S &=& S_g+S_\phi\,;
\cr
&&S_g=\int d^4 x \sqrt{- g}\frac{1}{2\kappa^2}R\,,
\cr
&&S_\phi=\int d^4 x \sqrt{- g} \left[W (X, \phi)
 - G (X, \phi) \Box \phi \right]\,;
\quad
 X \equiv - \frac{1}{2} g^{\mu \nu} \pa_\mu \phi \pa_\nu \phi\,,
 \label{action}
\end{eqnarray}
where $W$ and $G$ are arbitrary functions of $X$ and $\phi$.
For simplicity and definiteness, we consider Einstein gravity,
with $\kappa^2=8\pi G$,
but our discussion below can be easily extended to any metric
theory of gravity.
The energy momentum tensor of scalar field
is given by
 \begin{gather}
  T_{\mu \nu} = - 2 \frac{\delta S_\phi}{\delta g^{\mu \nu}} 
 = W_{, X} \pa_\mu \phi \pa_\nu \phi + W g_{\mu \nu} - G_{, X} \Box \phi
 \pa_\mu \phi \pa_\nu \phi
 + g_{\mu \nu} G_{, \rho} g^{\rho \sigma} \pa_\sigma \phi
 - (G_{, \mu} \pa_\nu \phi + G_{, \nu} \pa_\mu \phi)\,.
 \end{gather}

Our theory includes a Galileon scalar if 
$G\neq0$~\cite{galileon,g-inflation}.
If we consider the case,
\begin{gather}
 G (X, \phi) = 0\,,
\end{gather}
but retain a generic $W$, it corresponds to a K-essential
scalar \cite{k-inflation,garriga-mukhanov}, which includes a DBI model
 \cite{DBI,DBI-in-the-sky} as a special case.
If we consider the case,
\begin{gather}
 G (X, \phi) = 0\,,
\quad W (X, \phi) = X - V (\phi)\,, 
\end{gather}
we recover the conventional canonical scalar theory.

\section{Scalar field equation}
To derive the scalar field equation, we take the variation of
$S_\phi$ with respect to $\phi$,
\begin{eqnarray}
 \delta S &=& \int d^4 x \sqrt{- g} \left[ W_{, X} \delta X
 + W_{, \phi} \delta \phi - (G_{, X} \delta X + G_{, \phi} \delta \phi)
 \Box \phi - G \Box \delta \phi \right] 
\cr
 &=& \int d^4 x \sqrt{- g} \left[ (W_{, \phi} - G_{, \phi} \Box \phi) \delta \phi
 - (W_{, X} - G_{, X} \Box \phi) \nabla_\mu \delta \phi \nabla^\mu \phi
 + (G_{, X} \nabla^\mu X + G_{, \phi} \nabla^\mu \phi) \nabla_\mu \delta \phi
 \right]
\cr
 &=& \int d^4 x \sqrt{- g} \left[ (W_{, \phi} - G_{, \phi} \Box \phi)
 + \nabla_\mu \left[ (W_{, X} - G_{, X} \Box \phi) \nabla^\mu \phi \right]
 - \nabla_\mu (G_{, X} \nabla^\mu X + G_{, \phi} \nabla^\mu \phi) \right]
 \delta \phi\,,
\end{eqnarray}
where we have used the identity,
\begin{eqnarray}
 Q\,\delta X = - Q\,g^{\mu \nu} (\pa_\mu \delta\phi)\pa_\nu \phi
=-\partial_\mu(Q\,g^{\mu \nu} \delta\phi \pa_\nu\phi)+
\pa_\mu(Q\,g^{\mu \nu}\pa_\nu\phi)\delta\phi\,,
\end{eqnarray}
and dropped the surface term after integration by part.
Thus the field equation is given by
\begin{gather}
 \frac{1}{\sqrt{- g}} \pa_\mu \Bigl( \sqrt{- g} \Bigl[(W_{, X} - G_{, \phi})
 \nabla^\mu \phi - G_{, X} (\Box \phi \nabla^\mu \phi 
 + \nabla^\mu X) \Bigr] \Bigr) + W_{, \phi} - G_{, \phi} \Box \phi = 0\,.
\label{fieldeq}
\end{gather}

We introduce a unit timelike vector $n^\mu$ orthonormal to
 the $t=$ constant hypersurfaces,
\begin{gather}
  n^\mu = \frac{1}{\alpha} (1, 0, 0, 0) \,.
\end{gather}
For this vector, the expansion $K$ is 
given by\footnote{Note the change of notation from \cite{shibata-sasaki},
in which $K$ is defined by the `minus' of the expansion; $K=-n^\mu{}_{;\mu}$.}
\begin{gather}
 K \equiv n^\mu{}_{; \mu}
 = \frac{1}{\sqrt{- g}} \pa_\mu (\sqrt{- g} n^\mu)
 = \frac{1}{\sqrt{\gamma}} \pa_\tau \sqrt{\gamma}\,.
\end{gather}
where $\tau$ is the proper time of an observer with 
the 4-velocity $n^\mu$,
\begin{gather}
 d \tau = \alpha(t,x^k)\,dt\,.
\end{gather}

Since we concentrate on the superhorizon behaviour,
we neglect the spatial derivatives in (\ref{fieldeq}) to obtain
\begin{gather}
 - \frac{1}{\sqrt{\gamma}} \pa_\tau \Bigl( \sqrt{\gamma}
 \Bigl[(W_{, X} - G_{, \phi})
 \pa_\tau \phi - G_{, X} (\Box \phi \pa_\tau \phi + \pa_\tau X) \Bigr]\Bigr)
 + W_{, \phi} - G_{, \phi} \Box \phi = 0\,,
\end{gather}
where
\begin{eqnarray}
\Box\phi=-\frac{1}{\sqrt{\gamma}}\pa_\tau(\sqrt{\gamma}\pa_\tau\phi)
=-\pa_\tau^2\phi-K\pa_\tau\phi\,.
\end{eqnarray}
Then the above can be rewritten as
\begin{eqnarray}
 &&\Bigl( W_{, X} - 2 G_{, \phi}
 + 2 G_{, X} K \pa_\tau \phi \Bigr)
 \pa_\tau^2 \phi
 + (\pa_\tau G_{, X}) K(\pa_\tau \phi)^2 
 + G_{, X} (\pa_\tau K+K^2)(\pa_\tau \phi)^2 
\cr
\cr
&&\qquad
 + \Bigl[\pa_\tau (W_{, X} - G_{, \phi}) + K(W_{, X} - 2 G_{, \phi})\Bigr]
 \pa_\tau \phi - W_{, \phi} = 0 \,.
\label{Feq}
\end{eqnarray}

Here an important comment is in order.
In the Galileon case where $G_{, X} \neq 0$,
 the scalar field equation contains the second time derivative of the 
metric, $\pa_\tau K+K^2=\pa_\tau^2\sqrt{\gamma}/\sqrt{\gamma}$.
 As we shall see below, 
the conservation of the nonlinear curvature perturbation
can be proved only if the field equation contains only first time
derivatives of the metric. Thus one has to use the gravitational
field equations (the Einstein equations in the present case) to
eliminate the second or higher derivative terms.

The relevant components of the Einstein equations are
\begin{eqnarray}
 K^2 = 3 \kappa^2 E \,,\quad
 \pa_\tau K = - \frac{3}{2} \kappa^2 (E + W
 + g^{\mu \nu} G_{, \mu} \phi_{, \nu})
 = - \frac{3}{2} \kappa^2 (E + W - \pa_\tau G \pa_\tau \phi)\,,
\end{eqnarray}
or
\begin{eqnarray}
\pa_\tau K+K^2=\frac{3}{2}\kappa^2\left(E-W
 + \pa_\tau G \pa_\tau \phi \right)\,,
\label{2ndK}
\end{eqnarray}
where $E=T_{\mu\nu}n^\mu n^\nu$. Keeping only the time derivatives,
$E$ is given by
\begin{eqnarray}
E = \frac{1}{\alpha^2} T_{0 0}
=\Bigl[W_{,X}-G_{,X}(\Box\phi) \Bigr] (\pa_\tau\phi)^2
 - \pa_\tau G \pa_\tau \phi - W \,.
\end{eqnarray}
This gives
\begin{eqnarray}
\pa_\tau K + K^2 
&=& \frac{3}{2} \kappa^2 \left( \Bigl[ W_{,X}
 - G_{,X} (\Box \phi) \Bigr] (\pa_\tau\phi)^2 - 2 W \right)
\cr\cr
& =& \frac{3}{2} \kappa^2 \left[ \Bigl( W_{,X} + G_{,X} K \pa_\tau \phi
 + G_{,X} \pa_\tau^2 \phi \Bigr) (\pa_\tau\phi)^2 - 2 W \right] \,.
\end{eqnarray}

Eliminating $\pa_\tau K+K^2$ in (\ref{Feq}) by using
Eq.~(\ref{2ndK}), we obtain the scalar field equation involving
only first time derivatives of the metric,
\begin{eqnarray}
&& (W_{, X} - 2 G_{, \phi} + 2 G_{, X} K \pa_\tau \phi) \pa_\tau^2 \phi
 + (G_{, X X} \pa_\tau^2 \phi + G_{, X \phi}) K (\pa_\tau \phi)^3 
\cr\cr
&&\quad + \Bigl[(W_{, X X} - G_{, \phi X})\pa_\tau^2 \phi + W_{, X \phi}
 - G_{, \phi \phi}\Bigr](\pa_\tau \phi)^2 
+ K (W_{, X} - 2 G_{, \phi}) \pa_\tau \phi
 - W_{, \phi}
\cr\cr
&&\quad + \frac{3}{2} \kappa^2 G_{, X} (\pa_\tau \phi)^2 
\Bigl[ (W_{, X} + G_{, X} K \pa_\tau \phi + G_{, X} \pa_\tau^2 \phi)
 (\pa_\tau \phi)^2 - 2W \Bigr] 
 = 0 \,.
 \label{eom}
\end{eqnarray}

\section{Conservation law}

Quite generally a conservation law corresponds to
an integral of motion. This implies that it will be necessary
for the scalar field equation to be effectively first order in
time derivatives in order to derive a conservation law.
In the present case, 
to show the conservation of the nonlinear curvature perturbation,
we assume that the system has evolved into an attractor stage
so that the time derivative of the scalar field has become
a function of $\phi$,
\begin{gather}
 \pa_\tau \phi = f (\phi)\,.
\label{attractor}
\end{gather}
Note that this may be regarded as a generalization of the slow-roll case.
In this regime, the functions $G$ and $W$ become functions of $\phi$ only:
\begin{gather}
  G = G (X, \phi) = G \left(\frac{1}{2} f^2, \phi \right)\,,
\quad 
 W = W (X, \phi) = W \left(\frac{1}{2} f^2, \phi \right)\,.
\end{gather}
We also assume $f\neq0$. This implies that $\phi$ can be used
to determine the time slicing if desired.

We can rewrite Eq.~(\ref{eom}) as
\begin{eqnarray}
&& (W_{, X} - 2 G_{, \phi} + 2 G_{, X} K f) f_{, \phi} f
 + (G_{, X X} f_{, \phi} f + G_{, X \phi}) K f^3 
\cr
\cr
&&\quad
 + \Bigl[(W_{, X X} - G_{, \phi X})  f_{, \phi} f
+ W_{, X \phi} - G_{, \phi \phi}\Bigr] f^2
 + K (W_{, X} - 2 G_{, \phi}) f - W_{, \phi}
\cr
\cr
&&\quad
 + \frac{3}{2} \kappa^2 G_{, X} f^2 \Bigl[ (W_{, X} + G_{, X} K f
 + G_{, X} f_{, \phi} f) f^2 - 2W \Bigr]
 = 0\,.  
\end{eqnarray}
This equation can be arranged in the form,
$A (\phi) + K B (\phi) = 0$, or
\begin{gather}
-K=\frac{A(\phi)}{B(\phi)}\,, 
\label{evo}  
\end{gather}
 where $A$ and $B$ are given by
\begin{eqnarray}
 A (\phi) &=&
 (W_{, X} - 2 G_{, \phi}) f_{, \phi} f
+ \Bigl[(W_{, X X} - G_{, \phi X}) f_{, \phi}f
 + (W_{, X \phi} - G_{, \phi \phi})\Bigr] f^2
 - W_{, \phi} 
\cr
\cr
 &&
+ \frac{3}{2} \kappa^2 G_{, X} f^2 \Bigl[ (W_{, X}
 + G_{, X} f_{, \phi} f) f^2 - 2 W \Bigr] \,,
\\
 B (\phi) &=&
 2 G_{, X} f^2 f_{, \phi} + (G_{, X X} f_{, \phi} f + G_{, X \phi}) f^3
 + (W_{, X} - 2 G_{, \phi}) f
 + \frac{3}{2} \kappa^2 (G_{, X})^2 f^5 \,.
\end{eqnarray}

Now we recall that $K$ 
is expressed in terms of the metric components as
\begin{eqnarray}
K=\frac{\pa_\tau\sqrt{\gamma}}{\sqrt{\gamma}}
=\frac{3}{\alpha}\left(H+\dot{\psi}\right)\,,
\end{eqnarray}
where $H\equiv \dot a/a$ and $\dot{}=\pa/\pa t$.
Integrating $K$ along the integral curve of $n^\mu$,
that is, along $x^k=$ constant, from $t_\ma{i}$ to $t$, we obtain
\begin{eqnarray}
\int_{t_\ma{i}}^t dt'\alpha K
=3\left[ \ln \left( \frac{a (t)}{a (t_\ma{i})} \right) +  
\psi (t, x^k) - \psi (t_\ma{i}, x^k) \right]\,.
\label{Kint}
\end{eqnarray}
On the other hand, the integral of the right-hand side of
Eq.~(\ref{evo}) gives
\begin{eqnarray}
\int_{t_\ma{i}}^tdt'\alpha\frac{A(\phi)}{B(\phi)}
= \int_{\phi_\ma{i}}^\phi d \phi' \frac{A (\phi')}{f (\phi') B (\phi')}
 = F (\phi) - F (\phi_\ma{i})\,.
\label{phiint}
\end{eqnarray}
Therefore, combining Eqs.~(\ref{Kint}) and (\ref{phiint}), we find
\begin{eqnarray}
-3\left[ \ln \left( \frac{a (t)}{a (t_\ma{i})} \right) +  
\psi (t, x^k) - \psi (t_\ma{i}, x^k) \right]
= F (\phi) - F (\phi_\ma{i})\,,
\label{inteq}
\end{eqnarray}
for any $t$.

So far, we have not specified the time slicing.
Now let us choose the uniform $\phi$ slicing, 
$\phi\bigl(\tau(t,x^k),x^k\bigr)=\phi(t)$,
or regard $\phi$ as a time coordinate.
That is, we choose the comoving slicing where $n_\mu T^\mu{}_i=0$.
In this case, the equation $\phi(t)$ satisfies becomes
identical to the one for the homogeneous and isotropic universe,
and the fiducial scale factor $a(t)$ can be chosen to be the one
for this homogeneous and isotropic universe.

Here it is worth mentioning another particular nature of the 
Galileon field. An explicit expression for $n_\mu T^\mu{}_i$
at lowest order in the gradient expansion is
 \begin{gather}
 n_\mu T^\mu{}_i = \pa_\tau \phi \Bigl[ \bigl(W_{,X}
 + K G_{,X}\pa_\tau \phi - 2 G_{, \phi}\bigr)\partial_i\phi 
 - G_{,X} \pa_\tau \phi \,\pa_i(\pa_\tau \phi)\Bigr]\,.
\label{T0i}
 \end{gather}
Because of the presence of the term proportional
to $\pa_i(\pa_\tau \phi)$, it is clear that the uniform $\phi$
slicing does not necessarily
coincide with the comoving slicing in general. 
However, in the present case, we have assumed that the system
is in an attractor regime where $\pa_\tau \phi$ has become
a function of $\phi$ alone, as given by Eq.~(\ref{attractor}).
Therefore we have $\pa_i(\pa_\tau \phi)=\pa_i f=f_{,\phi}\pa_i\phi$.
That is, in the Galileon case, the uniform $\phi$ slicing coincides
with the comoving slicing provided that the sytem is in an attractor
regime.

Then Eq.~(\ref{inteq}) implies
\begin{eqnarray}
-3\ln \left( \frac{a (t)}{a (t_\ma{i})} \right)
= F (\phi) - F (\phi_\ma{i})\,,
\quad
\psi_c (t, x^k) = \psi_c (t_\ma{i}, x^k)\,,
\end{eqnarray}
where $\psi_c$ is $\psi$ evaluated on comoving slices.
This is a proof of the conservation of the nonlinear curvature perturbation
on comoving slices.
The key for the proof
is the attractor behaviour of the scalar field, Eq.~(\ref{attractor}).

\section{conclusion}
We have shown that the nonlinear curvature perturbation on 
comoving slices is conserved on superhorizon scales for
 a very general class of single-field inflation. 
It can be derived by using only the scalar field equation
if it contains only first derivatives of the metric, while
the gravitational equations are necessary if it contains second
or higher derivatives of the metric.
The key, necessary condition is that the scalar field 
is in an attractor regime
so that $\phi$ can be taken as a time coordinate.

\acknowledgements
This work is supported in part by Monbukagaku-sho 
Grant-in-Aid for the Global COE programs, 
hThe Next Generation of Physics, Spun from Universality 
and Emergenceh at Kyoto University.
The work of MS is supported by JSPS Grant-in-Aid for Scientific Research 
(A) No.~21244033,
and by Grant-in-Aid for Creative Scientific Research No.~19GS0219.
AN is supported by Grant-in-Aid for JSPS Fellows No.~21-1899.


\begin{thebibliography}{}

\bibitem{wmap-five}
WMAP Collaboration, E.~Komatsu {\em et~al.},
\newblock Astrophys.J.Suppl. {\bf 180}, 330 (2009), arXiv:0803.0547.

\bibitem{komatsu-spergel}
E.~Komatsu and D.~N. Spergel,
\newblock Phys. Rev. {\bf D63}, 063002 (2001), arXiv:astro-ph/0005036.

\bibitem{sasaki-wands}
M.~Sasaki and D.~Wands,
\newblock Classical and Quantum Gravity {\bf 27}, 120301 (2010).

\bibitem{maldacena}
J.~M. Maldacena,
\newblock JHEP {\bf 05}, 013 (2003), arXiv:astro-ph/0210603.

\bibitem{malik-wands}
K.~A. Malik and D.~Wands,
\newblock Class. Quant. Grav. {\bf 21}, L65 (2004), arXiv:astro-ph/0307055.

\bibitem{seery-lidsey}
D.~Seery and J.~E. Lidsey,
\newblock JCAP {\bf 0506}, 003 (2005), arXiv:astro-ph/0503692.

\bibitem{not-so-short}
K.~A. Malik,
\newblock JCAP {\bf 0703}, 004 (2007), arXiv:astro-ph/0610864.

\bibitem{russia-lifshitz}
E.~M. Lifshitz and I.~M. Khalatnikov,
\newblock Adv. Phys. {\bf 12}, 185 (1963).

\bibitem{russia-belinsky}
V.~a. Belinsky, I.~m. Khalatnikov, and E.~m. Lifshitz,
\newblock Adv. Phys. {\bf 31}, 639 (1982).

\bibitem{starobinsky}
A.~A. Starobinsky,
\newblock JETP Lett. {\bf 42}, 152 (1985).

\bibitem{bardeen}
J.~M. Bardeen,
\newblock Phys. Rev. {\bf D22}, 1882 (1980).

\bibitem{salopek-bond}
D.~S. Salopek and J.~R. Bond,
\newblock Phys. Rev. {\bf D42}, 3936 (1990).

\bibitem{nathalie-langlois}
N.~Deruelle and D.~Langlois,
\newblock Phys. Rev. {\bf D52}, 2007 (1995), arXiv:gr-qc/9411040.

\bibitem{sasaki-stewart}
M.~Sasaki and E.~D. Stewart,
\newblock Prog. Theor. Phys. {\bf 95}, 71 (1996), arXiv:astro-ph/9507001.

\bibitem{nambu-taruya}
Y.~Nambu and A.~Taruya,
\newblock Class. Quant. Grav. {\bf 13}, 705 (1996), arXiv:astro-ph/9411013.

\bibitem{sasaki-tanaka}
M.~Sasaki and T.~Tanaka,
\newblock Prog. Theor. Phys. {\bf 99}, 763 (1998), arXiv:gr-qc/9801017.

\bibitem{shibata-sasaki}
M.~Shibata and M.~Sasaki,
\newblock Phys. Rev. {\bf D60}, 084002 (1999), arXiv:gr-qc/9905064.

\bibitem{lyth-rodriguez}
D.~H. Lyth and Y.~Rodriguez,
\newblock Phys. Rev. {\bf D71}, 123508 (2005), arXiv:astro-ph/0502578.

\bibitem{rigopoulos-shellard1}
G.~I. Rigopoulos and E.~P.~S. Shellard,
\newblock Phys. Rev. {\bf D68}, 123518 (2003), arXiv:astro-ph/0306620.

\bibitem{rigopoulos-shellard2}
G.~I. Rigopoulos and E.~P.~S. Shellard,
\newblock JCAP {\bf 0510}, 006 (2005), arXiv:astro-ph/0405185.

\bibitem{yokoyama1}
S.~Yokoyama, T.~Suyama, and T.~Tanaka,
\newblock JCAP {\bf 0707}, 013 (2007), arXiv:0705.3178.

\bibitem{yokoyama2}
S.~Yokoyama, T.~Suyama, and T.~Tanaka,
\newblock Phys. Rev. {\bf D77}, 083511 (2008), arXiv:0711.2920.

\bibitem{yotanaka1}
Y.~Tanaka and M.~Sasaki,
\newblock Prog. Theor. Phys. {\bf 117}, 633 (2007), arXiv:gr-qc/0612191.

\bibitem{yotanaka2}
Y.~Tanaka and M.~Sasaki,
\newblock Prog. Theor. Phys. {\bf 118}, 455 (2007), arXiv:0706.0678.

\bibitem{takamizu-mukohyama}
Y.-i. Takamizu and S.~Mukohyama,
\newblock JCAP {\bf 0901}, 013 (2009), arXiv:0810.0746.

\bibitem{takamizu-sasaki}
Y.-i. Takamizu, S.~Mukohyama, M.~Sasaki, and Y.~Tanaka,
\newblock JCAP {\bf 1006}, 019 (2010), arXiv:1004.1870.

\bibitem{LMS}
D.~H. Lyth, K.~A. Malik, and M.~Sasaki,
\newblock JCAP {\bf 0505}, 004 (2005), arXiv:astro-ph/0411220.

\bibitem{afshordi}
N.~Afshordi and R.~H. Brandenberger,
\newblock Phys. Rev. {\bf D63}, 123505 (2001), arXiv:gr-qc/0011075.

\bibitem{separate-universe}
D.~Wands, K.~A. Malik, D.~H. Lyth, and A.~R. Liddle,
\newblock Phys. Rev. {\bf D62}, 043527 (2000), arXiv:astro-ph/0003278.

\bibitem{k-inflation}
C.~Armendariz-Picon, T.~Damour, and V.~F. Mukhanov,
\newblock Phys. Lett. {\bf B458}, 209 (1999), arXiv:hep-th/9904075.

\bibitem{garriga-mukhanov}
J.~Garriga and V.~F. Mukhanov,
\newblock Phys. Lett. {\bf B458}, 219 (1999), arXiv:hep-th/9904176.

\bibitem{DBI}
E.~Silverstein and D.~Tong,
\newblock Phys. Rev. {\bf D70}, 103505 (2004), arXiv:hep-th/0310221.

\bibitem{DBI-in-the-sky}
M.~Alishahiha, E.~Silverstein, and D.~Tong,
\newblock Phys. Rev. {\bf D70}, 123505 (2004), arXiv:hep-th/0404084.

\bibitem{g-inflation}
T.~Kobayashi, M.~Yamaguchi, and J.~Yokoyama,
\newblock Phys. Rev. Lett. {\bf 105}, 231302 (2010), arXiv:1008.0603.

\bibitem{galileon}
A.~Nicolis, R.~Rattazzi, and E.~Trincherini,
\newblock Phys. Rev. {\bf D79}, 064036 (2009), arXiv:0811.2197.

\bibitem{galileon-mizuno}
S.~Mizuno and K.~Koyama,
\newblock Phys. Rev. {\bf D82}, 103518 (2010), arXiv:1009.0677.

\bibitem{galileon-deffayet}
C.~Deffayet, O.~Pujolas, I.~Sawicki, and A.~Vikman,
\newblock JCAP {\bf 1010}, 026 (2010), arXiv:1008.0048.

\bibitem{KMD}
 K.~Kamada, T.~Kobayashi, M.~Yamaguchi and J.~Yokoyama,
\newblock arXiv:1012.4238 [astro-ph.CO].

\end{thebibliography}
\end{document}